\documentclass[conference]{IEEEtran}
\IEEEoverridecommandlockouts
\usepackage[cmex10]{amsmath}
\usepackage{eqparbox}
\usepackage[usenames, dvipsnames]{color}
\usepackage{mathtools,amssymb,bm,mathabx,adjustbox}
\usepackage{amstext}
\usepackage{amssymb}
\usepackage{graphicx}
\usepackage{color}
\usepackage{booktabs}
\usepackage{longtable}
\usepackage{multicol}
\usepackage{algorithmicx}
\usepackage[ruled]{algorithm}
\usepackage{algpseudocode}
\usepackage{algpascal}
\usepackage{algc}
\usepackage{amsfonts}
\usepackage{dsfont}
\usepackage{array}
\usepackage[most]{tcolorbox}
\usepackage{cases}
\usepackage{pgfplots}
\usepackage{caption}
\usepackage{subfigmat}
\usepackage{xr-hyper}
\usepackage{hyperref}
\usepackage{amsthm}
\DeclareCaptionLabelSeparator{periodspace}{.~}
\usepackage{float}
\usepackage{cite}
\usepackage{epstopdf}
\theoremstyle{remark}

\newcommand\ASTART{\bigskip\noindent\begin{minipage}[b]{0.5\linewidth}}
	
	\newcommand\AENDSKIP{\end{minipage}\bigskip}
\newcommand\AEND{\end{minipage}}
\ifCLASSOPTIONcaptionsoff
\usepackage[nomarkers]{endfloat}
\let\MYoriglatexcaption\caption
\renewcommand{\caption}[2][\relax]{\MYoriglatexcaption[#2]{#2}}
\fi
\theoremstyle{plain}
\newtheorem{thm}{\textbf{Theorem}}
\newtheorem{lem}{\textbf{Lemma}}

\theoremstyle{definition}

\newcommand*{\rom}[1]{\expandafter\@slowromancap\romannumeral #1@}

\newcommand{\RN}[1]{%
\textup{\uppercase\expandafter{\romannumeral#1}}%
}

\usepackage{standalone}
\graphicspath{{./Figures/}}
\usepackage{times}
\usepackage[T1]{fontenc}
\usepackage[english]{babel}
\usepackage{graphicx}
\usepackage{ amsfonts, dsfont,amssymb, graphicx, array, tabularx, booktabs,multicol,nccmath}
\usepackage{amsthm}
\usepackage{epsf,epsfig}
\usepackage{setspace}
\usepackage{subfigure}

\usepackage{hyperref}
\usepackage{multirow}
\usepackage{url}
\usepackage{color}
\usepackage[nolist,printonlyused]{acronym}      
\usepackage{comment}
\usepackage{cite}
\usepackage{bm}
\usepackage{tikz}
\usetikzlibrary{decorations.pathreplacing}

\usepackage{mathtools}

\usepackage{blindtext}

\usepackage{stfloats}

\newcommand{\gf}[1]{\textcolor{black}{{#1}}}

\newcommand{\mx}[1]{\mathbf{#1}}
\newcommand{\bs}[1]{\boldsymbol{#1}}

\definecolor{amber}{rgb}{1.0, 0.49, 0.0}
\definecolor{ao}{rgb}{0.0, 0.5, 0.0}

\def\R2#1{\textcolor{black}{#1}}
\def\R3#1{\textcolor{black}{#1}}

\renewcommand{\triangleq}{\mathbin{\setstackgap{S}{0pt}\stackMath\Shortstack{\smalltriangleup\\ =}}}
\usepackage[usestackEOL]{stackengine} 
\usepackage[utf8]{inputenc} 
\usepackage{url}
\usepackage{ifthen}
\usepackage{cite}

\usepackage{xcolor}
\def\BibTeX{{\rm B\kern-.05em{\sc i\kern-.025em b}\kern-.08em
    T\kern-.1667em\lower.7ex\hbox{E}\kern-.125emX}}
\begin{document}
\title{Optimal Transmitter Design and Pilot Spacing in MIMO Non-Stationary Aging Channels\\
\thanks{This work is supported in part by Digital Futures Project PERCy. G.Fodor was also supported by the Swedish Strategic Research (SSF) grant for the FUS21-0004 SAICOM project and the 6G-Multiband Wireless and Optical Signalling for Integrated Communications, Sensing and Localization (6G-MUSICAL) EU Horizon 2023 project, funded by the EU, Project ID: 101139176.}
}

\author{\IEEEauthorblockN{1\textsuperscript{st} Sajad ~Daei}
\IEEEauthorblockA{\textit{School of Electrical Engineering and Computer Science} \\
\textit{KTH Royal Institute of Technology, Stockholm, Sweden}\\
Stockholm, Sweden,\\
Email: sajado@kth.se}
\and
\IEEEauthorblockN{2\textsuperscript{nd} Gabor~Fodor}
\IEEEauthorblockA{\textit{School of Electrical Engineering and Computer Science} \\
\textit{KTH Royal Institute of Technology, Stockholm, Sweden}\\
\textit{ Ericsson Research, Stockholm, Sweden}\\
Stockholm, Sweden,\\
 Email: gaborf@kth.se}
\and
\IEEEauthorblockN{3\textsuperscript{rd} Mikael Skoglund}
\IEEEauthorblockA{\textit{School of Electrical Engineering and Computer Science} \\
\textit{KTH Royal Institute of Technology, Stockholm, Sweden}\\
Stockholm, Sweden,\\
Email: skoglund@kth.se}

}

\maketitle


\begin{abstract}

This work considers an uplink wireless communication system where multiple users with multiple antennas transmit data frames over dynamic channels. Previous studies have shown that multiple transmit and receive antennas can substantially enhance the sum-capacity of all users when the channel is known at the transmitter and in the case of uncorrelated transmit and receive antennas. However, spatial correlations stemming from close proximity of transmit antennas and channel variation between pilot and data time slots, known as channel aging, can substantially degrade the transmission rate if they are not properly into account. In this work, we provide an analytical framework to concurrently exploit both of these features. Specifically, we first propose a beamforming framework to capture spatial correlations. Then, based on random matrix theory tools, we introduce a deterministic expression that approximates the average sum-capacity of all users. Subsequently, we obtain the optimal values of pilot spacing and beamforming vectors upon maximizing this expression. Simulation results show the impacts of path loss, velocity of mobile users and Rician factor on the resulting sum-capacity and underscore the efficacy of our methodology compared to prior works.
\end{abstract}
\section{Introduction}
\gf{For the uplink of wireless multi-user}
communication systems, the integration of multiple transmit and receive antennas \gf{significantly enhances} transmission sum-capacity and spectral efficiency (SE). \gf{Several research results as well as large-scale deployments have demonstrated} the achievability of such improvements, especially in scenarios where  the channel state information is known at the transmitter and the channel characteristics of the uplink users remain uncorrelated at both the transmitter and receiver nodes \cite{rhee2004optimality,rusek2012scaling,jafar2001channel,jafar2004transmitter,lu2014overview,soysal2009optimality,hassibi2003much,Fodor:17}. Typically, channel estimation entails uplink users transmitting pilots to the base station (BS), enabling the BS to estimate the transmitted symbols of mobile users during data time slots based on the channels established during pilot time slots. However, in the context of mobile non-stationary wireless communications, where devices are in motion, the channels at pilot and data time slots undergo rapid fluctuations \cite{Truong:13,Fodor:23,non_stationary_model}. This phenomenon is known as channel aging. Neglecting the effects of channel aging can compromise the advantages of massive multiple-input multiple-output (MIMO) systems. Additionally, spatial correlations among transmit antennas, stemming from their close proximity, can adversely impact the sum-rate of all users. 
To address the challenges posed by antenna correlations, previous studies have explored optimal transmitter design under perfect and block-fading channel assumptions. These investigations encompass both scalar coding, represented by beamforming, and vector coding \cite{rusek2012scaling, jafar2004transmitter, soysal2009optimality, lu2014overview}. 
\gf{Driven by the demand for uplink data rates and coverage, and due to recent advances in antenna technologies, using multiple antennas and advanced precoding schemes at handheld user equipment devices have made uplink beamforming viable even in
compact small-sized mobile devices.} Furthermore, the optimality of multiple antenna transmissions in terms of the achievable uplink user bitrates has been extensively examined in various works, including \cite{jafar2004transmitter, rhee2004optimality}.
In addition to the aforementioned considerations, the mitigation of channel aging effects has been a central focus in various existing studies, as evidenced by key contributions such as \cite{Fodor:2021, Fodor:22, Abeida:10, Hijazi:10, Truong:13, Kong:2015, Yuan:20, Kim:20, Fodor:23, daei2024towards}. An initial investigation by \cite{Truong:13} \gf{has} delved into the impact of channel aging on the performance of single-input multiple-output (SIMO) systems, where uplink users operate with single antennas. \gf{That work has} proposed an autoregressive (AR) model for the time-varying channel, with the state transition matrix of the AR model being a scalar multiple of the identity matrix, effectively capturing time correlations. Building upon this foundation, \cite{Fodor:23} and \cite{daei2024towards} \gf{have} extended the approach to encompass a general state transition matrix. \gf{For fading stationary channels, which has exponentially decaying autocorrelation functions},\cite{Fodor:23} \gf{has} determined the optimal pilot spacing for a specific user by maximizing the deterministic equivalent SE. This deterministic expression, explored in works such as \cite{Hoydis:13, couillet2011deterministic, hachem2007deterministic}, provides accurate SE approximations when the number of receive antennas is sufficiently large. Additionally, reference \cite{daei2024towards} has delved into the explicit relationship between the velocity of mobile users and the correlation matrix in non-stationary environments, proposing a multi-frame optimization framework to identify optimal values for the number of frames, frame sizes, and pilot and data powers.

While the impact of spatial and temporal channel correlations has been distinctively studied in the SIMO case, this work proposes a framework to concurrently exploit both spatial and temporal correlations in the MIMO case. Firstly, we propose a beamforming framework to reshape the transmitted symbols of multi-antenna users, capturing both time and spatial channel correlations. Subsequently, utilizing tools from random matrix theory, we present a deterministic expression to approximate the expectation of sum-capacity of all uplink users. Further, we determine optimal values for beamforming vectors and pilot spacing by maximizing the proposed sum-capacity expression in Rician non-stationary MIMO systems. Numerical experiments show the effectiveness of the proposed approach and also the impacts of various factors, including Doppler frequency, Rician factors, and path loss, on the resulting sum-rate of all users and optimal transmitter design.
\textit{Notation}.
$\mx{I}_N$ represents the identity matrix of size $N$. The column stacking vector operator, denoted as $\textbf{vec}$, transforms a matrix $\mx{X}\in\mathbb{C}^{M\times N}$ into its vectorized version $\mx{x}\triangleq \textbf{vec}(\mx{X})\in\mathbb{C}^{M N\times 1}$. ${\rm vec}^{-1}(\cdot)$ reshapes a vector $\mx{x}$ into a matrix $\mx{X}$ by stacking the columns vertically.
The all-one vector of size $N$ is represented as $\mx{1}_N\in\mathbb{R}^{N\times 1}$. 
The $(k,l)$-th element of the matrix $\mx{A}$ is denoted as $[\mx{A}]_{k,l}$. The autocorrelation and autocovariance matrix of a random vector $\mx{z}$ are denoted by $\mx{R}_{\mx{z}}$ and $\mx{C}_{\mx{z}}$, respectively.

\section{System model}
\begin{figure}[htbp]
    \centering
\includegraphics[scale=.2]{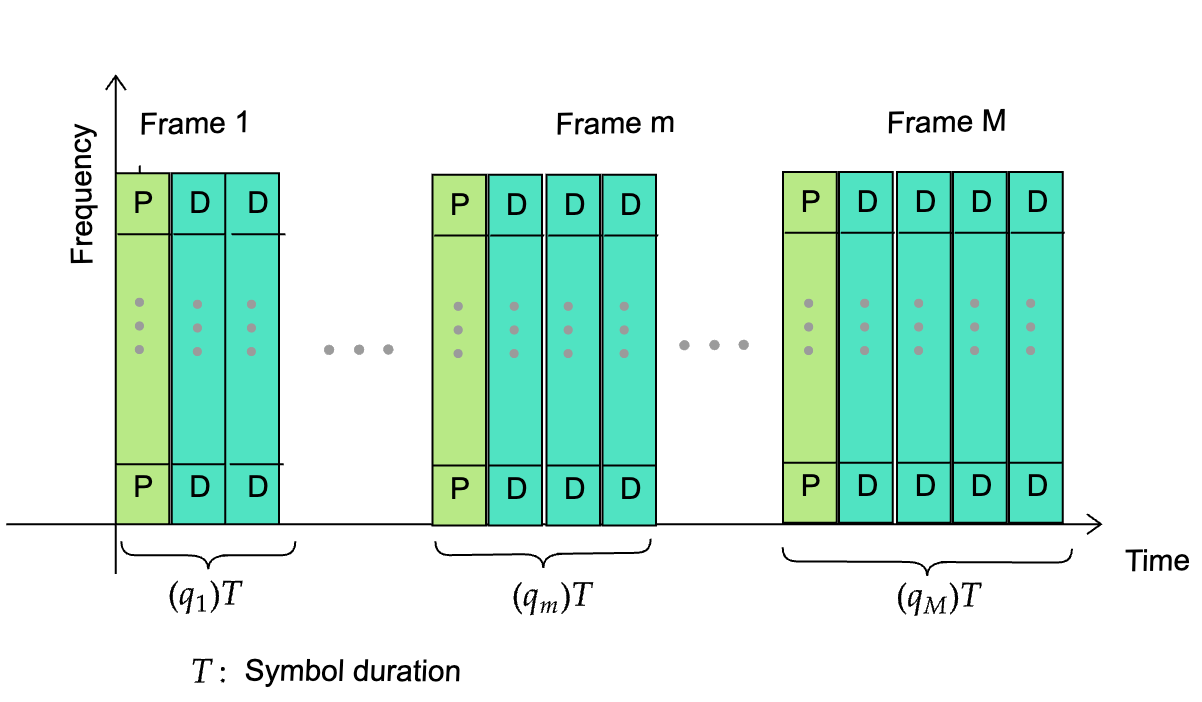}
    \caption{A diagram illustrating multi-frame data transmission for each user, where $q_1, ..., q_M$ represent the data length of frames $1, ..., M$. The initial time slot of each frame is designated for pilot transmission and contains $\tau_{\rm p}$ symbols in the frequency domain.}
    \label{fig:framework}
\end{figure}

We consider an uplink communication system where $K$ users with Doppler frequencies ${f_{d}}_1, ..., {f_{d}}_K$ send their data symbols in the form of $M$ frames with length $q_1,..., q_M$ (collected in a vector $\mx{q}\triangleq[q_1,..., q_M]^{\rm T}$) towards the serving \gf{BS} in the uplink as shown in Figure \ref{fig:framework}. The carrier frequency is shown by $f_c$ and the sampling time is assumed to be $T=1$ [time unit] without loss of generality throughout the paper. Each frame consists of one pilot time slot and the rest is data time slots. Each user is equipped with $N_t$ number of transmit antennas. The BS is equipped with $N_r$ number of receive antennas distributed as uniform linear array (ULA) with antenna distance $d$. The channel between each user and the BS at time slot $t$, is shown by the matrix $\mx{H}(t)\in\mathbb{C}^{N_r\times N_t}$ which is assumed to be non-stationary random process. Its vectorized version is denoted by $\mx{h}(t)\triangleq {\rm vec}(\mx{H}(t))\in\mathbb{C}^{N}$ where $N\triangleq N_t N_r$. Specifically, the mean and covariance of $\mx{h}(t)$ are denoted by $\overline{\mx{h}}(t)\in\mathbb{C}^{N\times 1}$ and $\mx{C}_{\mx{h}}\triangleq \mathds{E}[\widetilde{\mx{h}}(t)\widetilde{\mx{h}}^H(t)]$. Here, $\widetilde{\mx{h}}(t)\triangleq \mx{h}(t)-\overline{\mx{h}}(t)$ is the centered channel. Furthermore, the cross-covariance matrix of the channel is defined as $\mx{C}_{\mx{h}}(t_1,t_2)\triangleq \mathds{E}[\widetilde{\mx{h}}(t_1)\widetilde{\mx{h}}^H(t_2)]$.
At the current time $t_1$, the channel can be expressed as a sum of two components. The first component is associated with information exhibiting correlation with the \gf{preceding} time $t_2$, while the second component represents the innovative information introduced at time $t_1$. This relationship is provided as follows \cite[Prop. 1]{daei2024towards}:
{
\begin{align}
\label{eq:AR1}
&\widetilde{\mx{h}}(t_1)=\underbrace{ \mx{A}(t_1,t_2) \widetilde{\mx{h}}(t_2) }_{\textup{Correlation information}}+\underbrace{\bs{\xi}(t_1)}_{\textup{Innovative information}}.  
\end{align}}
Here, $\mx{A}(t_1,t_2)=\mx{C}(t_1,t_2)\mx{C}(t_2)^{-1}$ denotes the state transition matrix, and $\bs{\zeta}(t_1)$ is the error of representing the channel at time $t_1$ based on the channel at time $t_2$. The correlation matrix between times $t_1$ and $t_2$ is defined as $\mx{P}_{\mx{h}}(t_1,t_2)\triangleq \mx{C}_{\mx{h}}^{-\tfrac{1}{2}}(t_1)\mx{C}_{\mx{h}}(t_1,t_2)\mx{C}_{\mx{h}}^{-\tfrac{1}{2}}(t_2)$ and is explicitly related to the Doppler frequency \cite[Prop. 2]{daei2024towards}.

The observed measurements at pilot time slots corresponding to frames $m-1$, $m$ and $m+1$ and user $k$ can be collected in a matrix-vector form given by :
{
\begin{align}\label{eqn:measurements_pilot_vectorized}
   & {\mx{y}}_{{\rm p},k}=\alpha_k\sqrt{P_{{\rm p},k}}\widetilde{\mx{S}}\widetilde{\mx{h}}_{{\rm p},k}+\widetilde{\mx{e}}_k\in\mathbb{C}^{3N_r\tau_{\rm p} \times 1},
\end{align}}
where $\alpha_k$ is the path loss of the $k$-th user, ${\mx{h}_p}=\begin{bmatrix}
       \mx{h}_k(-q_{m-1}-1))\\
       \mx{h}_k(0)\\
       \mx{h}_k(q_{m}+1))
   \end{bmatrix}\in\mathbb{C}^{3 N\times 1}$ is the channel vector at pilot time slots corresponding to previous, current and subsequent frames, $\widetilde{\mx{S}}\in\mathbb{C}^{3N_r \tau_p\times 3N}$ is a matrix composing of pilots in the frequency domain corresponding to frames $m-1$, $m$ and $m+1$. $\tau_p$ is the number of pilots in the frequency domain. $P_{\rm p,k}=\tfrac{{P_{{\rm p},k}}_{\max}}{M}$, where ${P_{{\rm p},k}}_{\max}$ denotes the maximum (total) pilot power of the $k$-th user. At pilot time slots, we assume that the pilots are distributed in an orthogonal way in frequency domain as shown in Figure \ref{fig:framework} and there is no interference term at pilot times.
The measurements received at the BS side at data time slot $i$ of $m$-th frame are as follows:
{
\begin{align}
    &\mathbf{y}_{\rm d}(i)= \sum_{k=1}^K \alpha_{k} \mx{H}_k(i) \mx{x}_{k}(i)
+\mathbf{n}_{d}(i)\in\mathbb{C}^{N_r \times 1},
\label{eq:data_measurements}
\end{align}}
where $\sigma_{\rm d}^2$ is the variance of noise, $\mx{x}_{k}(i)\triangleq \mx{w}_k s_k(i)\in\mathbb{C}^{N_t\times 1}$ denotes the transmitted signal of user $k$
at time slot $i$ of $m$-th frame with transmit data power $P_{{\rm d},k}=\mathds{E}[|s_k(i)|^2]$. $s_k(i)\in\mathbb{C}$ is the transmitted symbol of user $k$
at time slot $i$. $\mx{w}_k\in\mathbb{C}^{N_t\times 1}$ is the beamforming vector of $k$-th user which forms a directional beam towards a specific direction.  
 
\section{Proposed approach}
We outline the steps of our analytical framework. Initially, we provide the covariance matrix of the linear minimum mean square error (LMMSE) estimate that estimates the channel considering temporal correlations with previous and next frames. Utilizing these channel estimates, we apply a minimum mean square error (MMSE) combiner at the receiver to estimate the symbol of of each user. Subsequently, we derive an expression for the instantaneous signal-to-interference-plus-noise ratio (SINR) per time slot for each user, from which we formulate the SE. As channel estimates introduce randomness, resulting in a random SE process, we employ concentration inequalities from random matrix theory to demonstrate that the instantaneous SE converges around a deterministic expression as $N_r$ becomes sufficiently large. The resultant SE expression depends on frame sizes $q_1,..., q_M$, $M$ and beamforming vectors, while being influenced by the temporal and spatial correlations of the channel. Finally, we determine optimal values for beamforming vectors, the number $M$ of frames, frame sizes based om maximizing the sum of SEs of all users, under some constraints on the beamforming vectors.

The resulting SE and SINR are dependent on the performance of the channel estimates, which is characterized by their covariance matrices. Therefore, we include the covariance matrix of LMMSE channel estimates below:
\begin{lem}
\label{lem:mmsechannel}	
The covariance matrix of the \textup{LMMSE} channel estimate at data time slot $i$ of frame $m$ corresponding to user $k$
based on the received measurements at pilot time slots of previous and next frames is provided by:
\begin{align}
\label{eq:rmmse}
\scalebox{.8}{$\mx{C}_{\widehat{\mx{h}}_k}(i)
= \mx{E}_{k,m}(\mx{q}, i) \left(\mx{M}_{k,m}(\mx{q})+\tfrac{\sigma^2_{{\rm p},k}}{\tau_{\rm p} \alpha_k^2 P_{{\rm p},k}}\mx{I}_{3N_r}\right)^{-1} \mx{E}_{k,m}^{\rm H}(\mx{q}, i),$}
\end{align}
where 
    \begin{align}
&\scalebox{.7}{$    \mx{E}_{k,m}(\mx{q},i)\triangleq
\begin{bmatrix}
    \mx{C}_{\mx{h}}(\delta_{m-2},i)&\mx{C}_{\mx{h}}(\delta_{m-1},i)&\mx{C}_{\mx{h}}(\delta_{m}, i) 
\end{bmatrix};$}\label{eqn:E_m}\\
& \mx{M}_{k,m}(\mx{q})\triangleq  \scalebox{.7}{${}\begin{bmatrix}
    \mx{C}_{\mx{h}}(\delta_{m-2})&\mx{C}_{\mx{h}}(\delta_{m-2},\delta_{m-1})&\mx{C}_{\mx{h}}(\delta_{m-2},\delta_{m})\\
    \mx{C}_{\mx{h}}(\delta_{m-1},\delta_{m-2})&\mx{C}_{\mx{h}}(\delta_{m-1})&\mx{C}_{\mx{h}}(\delta_{m-1},\delta_{m})\\
    \mx{C}_{\mx{h}}(\delta_{m}, \delta_{m-2})&\mx{C}_{\mx{h}}(\delta_{m},\delta_{m-1})&\mx{C}_{\mx{h}}(\delta_{m})
    \end{bmatrix}$}\label{eqn:M_m},
\end{align}    
and $\sigma_{{\rm p},k}^2$ is the variance of each element of $\widetilde{\mx{e}}_{{\rm p},k}$ and $\delta_m\triangleq\sum_{l=1}^m q_l +1$.
\end{lem}
Given the channel estimates above, we can now calculate the data estimates using the MMSE method.
We assume that the transmitted data vector of user $k$ has zero mean with variance $P_{d,k}$.
Specifically, the BS estimates the transmitted symbol of user $k$ at time slot $i$ based on channel estimates provided at $i$ which exploits the temporal correlations and is denoted by $\mx{z}_k\triangleq \mx{z}_k(i)\triangleq \widehat{\mx{h}}_k(i) $. Then, based on the channel estimate, the symbol estimate $\widehat{s}_k(i)=\mx{g}_k^\star(i)\mx{y}_{\rm d}$ where $\mx{g}_k^\star(i)\in\mathbb{C}^{1\times N_r}$ is the MMSE receiver to estimate the symbol of $k$-th user and is obtained as:
{
\begin{align}
\scalebox{.8}{$
   \mx{g}_k^{\star}(i)\triangleq \mathop{\arg\min}_{\mx{g}\in\mathbb{C}^{1\times N_r}} \mathds{E}_{s_k|\mx{z}_k(i)}|\mx{g}\mx{y}_{\rm d}-s_k(i)| =\alpha_k\sqrt{P_k}\mx{w}_k^H \mx{J}_k^H \mx{F}(i)^{-1}$}
\end{align}}
where 

\begin{align}
    &\mx{F}(i)\triangleq\sum_{k=1}^K \alpha_k^2\mathcal{A}_k(\mx{D}_k)+\sigma^2_d\mx{I}_{N_r}, \mx{D}_k\triangleq\mx{Q}_k+ \mx{z}_k\mx{z}_k^H,\\
    &\mx{J}_k\triangleq{\rm vec}^{-1}(\mx{z}_k)\in\mathbb{C}^{N_r\times N_t},\mx{Q}_k\triangleq\mx{C}_{\mx{h}_k(i)}- \mx{C}_{\widehat{\mx{h}}_k(i)},\\
    &\scalebox{.8}{$\mathcal{A}_k(\mx{D}_k)\triangleq\begin{bmatrix}
        \langle {\mx{D}^{\prime}}_k^{(1,1)},\mx{C}_{\mx{x}_k} \rangle& \hdots& \langle {\mx{D}^{\prime}}_k^{(1,N_r)},\mx{C}_{\mx{x}_k} \rangle\\  
        \vdots &\ddots&\vdots\\
         \langle {\mx{D}^{\prime}}_k^{(N_r,1)},\mx{C}_{\mx{x}_k} \rangle& \hdots& \langle {\mx{D}^{\prime}}_k^{(N_r,N_r)},\mx{C}_{\mx{x}_k} \rangle
    \end{bmatrix},$}\label{eq:A_operator}\\
    &{\mx{D}^{\prime}}_k\triangleq \mx{P}_{\rm c}^H \mx{D}_k \mx{P}_{\rm c},
\end{align}
where $\mx{P}_{\rm c}$ of size $N\times N$ is the commutation matrix which transforms the vectorized version of a matrix to the vectorized version of its transpose, $\mx{D}_k^{(i,j)}\in\mathbb{C}^{N_t\times N_t}$ is the $(i,j)$ block of matrix $\mx{D}_k$ and $\mx{C}_{\mx{x}_k}\triangleq P_{{\rm d},k} \mx{w}_k\mx{w}_k^H$.
The following theorem, whose proof is in Appendix \ref{proof.thm.instantanous_sinr}, calculates the instantaneous SINR for user $k$.
\begin{thm}\label{thm.Instantanous_SINR}
Let the channel estimate of the $k$-th user at time slot $i$ be $\mx{z}_k(i)$. Additionally, the receiver utilizes the {MMSE} combiner to estimate the data symbols of users during time slot $i$. Consequently, the instantaneous {SINR} for user $k$ at time slot $i$ is given by:
\begin{align}\label{eq:Instantanous_SINR_simplified}
     &\gamma_k(\mx{q},i,\mx{z}_k(i)= \alpha_k^2\mx{z}_k^{\rm H}\left(\mx{C}_{\mx{x}_k}\otimes \left(\mx{F}(i)- \alpha_k^2  \mx{z}_k \mx{z}_k^{\rm H}\right)^{-1} \right)\mx{z}_k\nonumber\\
     &~~~~~~~~~~~~~\triangleq\alpha_k^2  \mx{z}_k^{\rm H} \left( \mx{C}_{\mx{x}_k} \otimes \mx{F}_k^{-1}(i)\right) \mx{z}_k,
     \end{align}
     where $\mx{F}_k\triangleq \mx{F}_k(i)\triangleq \mx{F}(i)-\alpha_k^2 \mx{z}_k\mx{z}_k^{\rm H}
$.
\end{thm}

The latter result immediately leads to finding the random {SE} of user $k$ as follows:
\begin{align}\label{eq:random_SE}
\textup{SE}_k\Big(\mx{q},i,
\mx{z}_k(i)\Big)\triangleq\log\Big(1+\gamma_k(\mx{q},i,\mx{z}_k(i))\Big),
\end{align}
which is a random variable. In the following theorem whose proof is provided in Appendix \ref{proof.thm.stiel}, utilizing concentration inequality findings from the tools of random matrix theory, as provided in \cite{bai2010spectral}, we provide a deterministic equivalent formulation for the SE. This expression offers a suitable approximation for the average SE in scenarios where the number of BS antennas is sufficiently large.

\begin{thm}\label{thm.stiel}
 Define $\mx{S}\triangleq \sum_{k=1}^K \alpha_k^2\mx{Q}_k$ and $\rho_{\rm d}\triangleq \sigma_{\rm d}^2$. Let $\mx{W}\triangleq[\mx{w}_1,..., \mx{w}_K]\in\mathbb{C}^{N_t\times K}$ be the beamforming matrix. When $N_r$ is sufficiently large ($N_r\rightarrow \infty$), the instantaneous {SE} in \eqref{eq:random_SE} at time slot $i$ for user $k$ is concentrated around a deterministic expression given by:
\begin{align}\label{eq:DSE}
  &\scalebox{.7}{${\textup{SE}}_k^{\circ}(\mx{q}, i,\mx{W})\triangleq\log\Bigg(1+P_{{\rm d},k}\left\langle \mx{R}_{\mx{z}_k}, \mx{w}_k\mx{w}_k^H\otimes \left(\sum_{\substack{l=1\\ l\neq k}}^K \tfrac{\alpha_l^2\mathcal{A}_l(\mx{R}_{\mx{z}_l})}{1+\alpha_l^2\omega_l}+\mx{S}+\rho_{\rm d} \mx{I}_{N_r}\right)^{-1} \right\rangle \Bigg)$},
\end{align}
where $\omega_j, j=1,..., K, j\neq k$ are the solution of the following system of equations:
\begin{align}\label{eq:fixed}
   &\scalebox{.9}{$\omega_j =\left\langle \mathcal{A}_j( \mx{R}_{\mx{z}_j}),\left(\sum_{\substack{l=1\\l\neq j}}^K\tfrac{\alpha_l^2\mathcal{A}_l(\mx{R}_{\mx{z}_l})}{1+\alpha_l^2\omega_l}+\mx{S}+\rho_{\rm d} \mx{I}_{N_r}\right)^{-1}\right\rangle$}.
\end{align}
\end{thm}
\noindent The deterministic equivalent SE provided in \eqref{eq:DSE} depends on the beamforming vectors $\mx{w}_1,..., \mx{w}_K$ of size $N_t\times 1$. 
In what follows, we provide an optimization problem that finds the optimal values of $M$, $\mx{q}$, and beamforming vectors :
\begin{align}\label{eq:opt_problem}
&\scalebox{.9}{$\max_{\mx{q},M,\mx{W}} \tfrac{\sum_{k=1}^K\sum_{l=1}^{\delta_M -1}\text{SE}_k^{\circ}(\mx{q}, l,\mx{W})}{\delta_M -1},
~~{\rm s.t.}~\|\mx{w}_i\|_2=1, i=1,..., K,$}
\end{align}
in which the objective function is obtained by taking the average of SEs over all time slots of all frames corresponding to all users and is denoted by SSE in simulations.
The latter optimization problem can be solved using standard numerical optimization techniques. Note that the beamforming matrix $\mx{W}$ is fixed over all time slots of all frames.

\section{Numerical Results}
\label{Sec:simulations}
\begin{figure}[htbp]
    \centering
    \includegraphics[scale=.2,trim={0cm 0cm 0cm 0cm}]{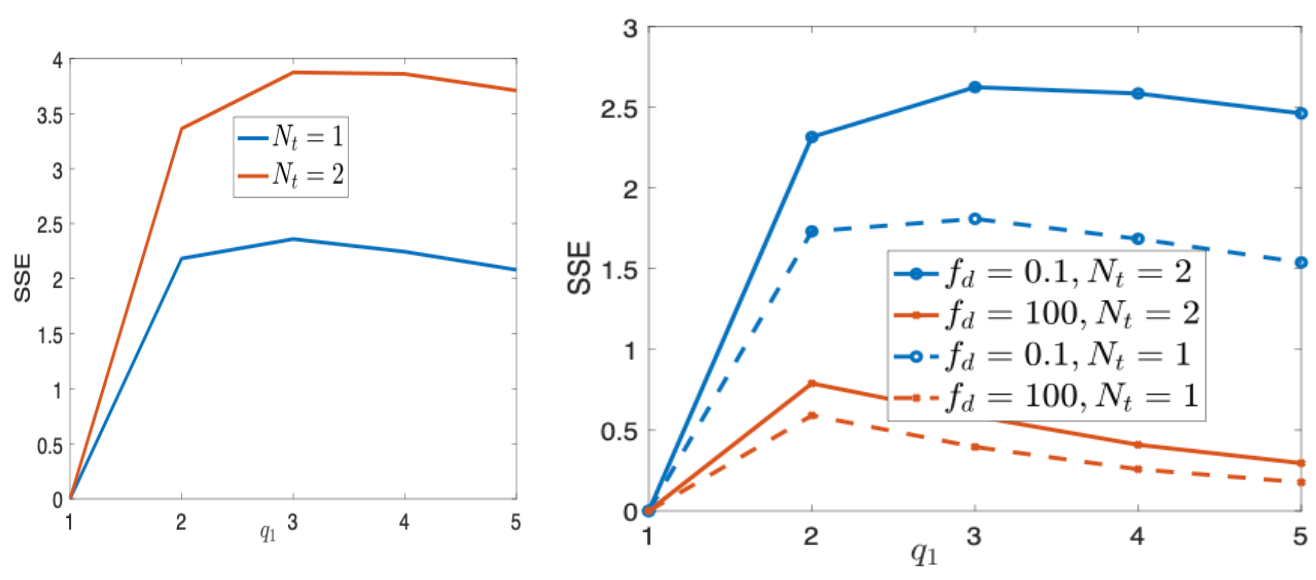}
    \caption{Left Image: SSE versus number of transmit antennas. Right image: SSE versus Doppler frequency.  The used parameters are $K=3, f_c=1000, PL_i=0, q_{\max}=5, M=1, N_r=10, {K_f}_i=0, {{{P_{{\rm p},i}}}}_{\max}={{P_{{\rm d},i}}}_{\max}=1, \sigma_d^2=0.01,  N_r=10, \tau_p=2, \rho_{\rm T}=0.9, \rho_{\rm R}=0$, with the exception of ${f_d}_i=0.1$ in the left image.}
    \label{fig:isit1}
\end{figure}

\begin{figure}[htbp]
    \centering
    \includegraphics[scale=.2,trim={0cm 0cm 0cm 0cm}]{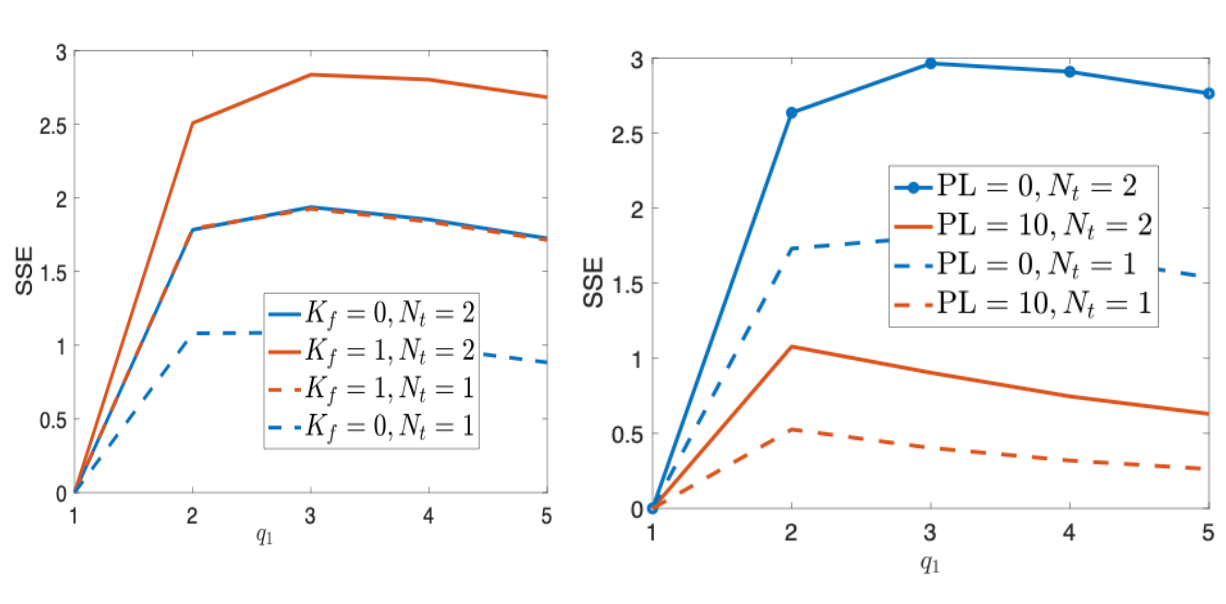}
    \caption{Left image: {SSE} versus Rician Factor. Right image: SSE versus path loss. The used parameters are $K=3, f_c=1000, {f_d}_i=0.1, PL_i=0, q_{\max}=5, M=1, N_r=10, {{{P_{{\rm p},i}}}}_{\max}={{P_{{{\rm d},i}}}}_{\max}=1, \sigma_d^2=0.01,  N_r=10, \tau_p=2, \rho_{\rm T}=0.9, \rho_{\rm R}=0$, with the exception of ${\rm PL}_i=0$ in the left image and ${K_f}_i=0$ in the right image.}
    \label{fig:isit2}
\end{figure}

In this section, we examine numerical experiments to evaluate the performance of our proposed method. The Doppler frequencies, pilot power, data power, Rician $K$ factor, path loss and channel variances are shown respectively by ${f_d}_i, {{P_{{\rm p,i}}}}_{\max}, {{P_{{\rm d},i}}}_{\max}, {K_f}_i, {\rm PL}_i, \sigma_{\mx{h}_i}^2$ where $i$ corresponds to user $i$ and ${\rm PL}_i=20\log(\alpha_i)$.
We consider a Kronecker model for the correlation matrix $\mx{P}(t_1,t_2)$ . For the transmit channel, we consider the transmit correlation matrix to be $\rho_{\rm T} \mx{1}_{N_t}\mx{1}_{N_t}^T+(1-\rho_{\rm T})\mx{I}_{N_t}$ with $\rho_{\rm T}=0.9$. Similarly, for the receive channel, we consider $\rho_{\rm R}=0$. 
In the first experiment shown in Figure \ref{fig:isit1}, we investigated the impact of multiple transmit antenna and Doppler on the optimal frame size $q$ in single frame scenario $M=1$. The left image of Figure \ref{fig:isit1} shows the effect of having multiple transmit on the the sum of spectral efficiencies (denoted by SSE) of all users. While increasing the number of transmit antennas improves the sum-capacity, it does not necessarily change the optimal frame size with the same settings. The right image of Figure \ref{fig:isit1} shows that having higher Doppler frequencies of mobile users leads to smaller optimal frame sizes. The improvement of multiple transmit antennas over single antenna is more enhanced in lower Doppler frequencies. In the left image of Figure \ref{fig:isit2}, the effects of Rician $K$ factor is investigated which indeed measures the strength of line of sight (LoS) component relative to the non-LoS components. We observe that the benefits of large $K$ factor with single transmit antenna and multiple transmit antenna and with zero $K$ factor can be comparable in the settings provided in the caption of Figure \ref{fig:isit2}. Larger $K$ factor shifts the optimal frame size to the right, suggesting using more data time slots in both single and multi-antenna cases. The right image of Figure \ref{fig:isit2} suggests using less data time slots in the case of large path loss of the channels. Moreover, the improvement of exploiting multiple transmit antennas with the proposed strategy is more enhanced in low path loss conditions.

\section{Conclusion} 
This study delved into uplink communication systems operating in fast-fading Rician non-stationary channels, taking into account the aging process between consecutive pilot time slots. The spatial correlations of transmit antennas and the impact of channel aging pose challenges to the benefits of massive MIMO. In response, we proposed a framework that leverages both spatial and temporal correlations. Specifically, we introduced an alternative expression for the sum-capacity in the multiple-antenna scenario. Subsequently, we optimized transmitter parameters, including beamforming, frame size, and the number of frames, by maximizing this expression. Our findings indicate that both the Rician factor and the number of transmit antennas contribute to enhancing the sum-capacity. Numerical experiments substantiated the efficacy of our approach in optimizing frame-related parameters across diverse scenarios, underscoring its practical significance. Notably, we observed that deploying multiple transmit antennas at mobile users can compensate for the absence of Line-of-Sight (LoS) channel components.
\appendices
\section{Proof of Theorem \ref{thm.Instantanous_SINR}}\label{proof.thm.instantanous_sinr}
Initially, we calculate the power of the estimated symbol $\mathds{E}|\widehat{s}_k(i)|^2$ and subsequently disentangle terms associated with user $k$ from interference terms. Following this, we determine the SINR by dividing the signal power by the sum of interference and noise power. More explicitly, 
\begin{align}
   &  \mathds{E}_{\mx{n}_d,\mx{h}_k(i)|\mx{z}_k(i)}  |\widehat{s}_k(i)|^2=
   \alpha_k^2 \left\langle \mathcal{A}_k(\mx{z}_k\mx{z}_k^{\rm H}), \mx{g}_k^{\rm H}\mx{g}_k \right\rangle+\nonumber\\
   &\scalebox{.9}{$\sum_{\substack{l=1\\l\neq k}}^K\alpha_l^2
   \left\langle \mathcal{A}_l(\mx{z}_k\mx{z}_k^{\rm H}), \mx{g}_k^{\rm H} \mx{g}_k \right\rangle+\sum_{l=1}^K \alpha_l^2 \left\langle \mathcal{A}_l(\mx{Q}_l),  \mx{g}_l^{\rm H} \mx{g}_l \right\rangle+\sigma^2_d \|\mx{g}_k\|_2^2.$}
\end{align}
The SINR can be written as $\gamma_k(\mx{q}_k,i,\mx{z}_k(i))=\tfrac{\left\langle \mx{g}_k^{\rm H} \mx{g}_k, \mx{F}-\mx{F}_k \right\rangle}{\left\langle \mx{g}_k^{\rm H} \mx{g}_k, \mx{F}_k \right\rangle}.$
    %
By assuming that the covariance matrix of the transmitted signal is unit-rank, the fact that the adjoint operator $\mathcal{A}^{\rm Adj}(\mx{Z})=\mx{C}_{\mx{x}_k}\otimes \mx{Z}$ for arbitrary matrix $\mx{Z}$ and appling the Woodbury matrix lemma \cite{horn2013matrix}, the result \eqref{eq:Instantanous_SINR_simplified} is achieved.
\section{Proof of Theorem \ref{thm.stiel}}\label{proof.thm.stiel}
First, we use \eqref{eq:Instantanous_SINR_simplified} and take expectation over $\mx{z}_k$ to find the expected SINR of user $k$. The resulting expression is the Frobinous inner product between a known matrix and the Stieltjes transform a finite measure.

Define the expectation of SE and SINR of user $k$,  by
\begin{align}
&\scalebox{.9}{$\overline{\textup{SE}}_k(\mx{q},i,\mx{W})\triangleq \mathds{E}\left[\textup{SE}_k(\mx{q},i,
\mx{z}_k(i))\right], \overline{\gamma}_k(\mx{q},i,\mx{W})\triangleq  \mathds{E}\left[\gamma_k(\mx{q},i,\mx{z}(i))\right]$}.
\end{align}
The aim is to find an equivalent deterministic expression denoted by ${\rm SE}_k^{\circ}(\mx{q},i, \mx{W})\triangleq \log(1+\gamma_k^\circ(\mx{q},i,\mx{W}))$ that approximates the average SE $\overline{\textup{SE}}_k(\mx{q},i,\mx{W})$ at time slot $i$. Here, $\gamma_k^\circ(\mx{q},i,\mx{W})$ is a deterministic equivalent expression for the SINR to be defined later. First, we claim that $\overline{\textup{SE}}_k(\mx{q},i,\mx{W})$ is close to the equivalent deterministic SE expression as long as $\overline{\gamma}(\mx{q},i,\mx{W})$ is sufficiently close to $\gamma_k^\circ(\mx{q},i,\mx{W})$ as is shown in the following error bound for Jensen's inequality \cite{jensen_gap}:
\begin{align}\label{eq:jensen}
 &   \scalebox{.78}{$ \tfrac{1}{N_r}\left|\overline{\textup{SE}}_k(\mx{q},i,\mx{W})-\log(1+{\gamma}_k^{\circ}(\mx{q},i,\mx{W}) )\right|\le\tfrac{c^\prime}{N_r}\left[\Big|\overline{\gamma}_k(\mx{q},i,\mx{W})-{\gamma}_k^{\circ}(\mx{q},i,\mx{W}) \Big|\right],$}
\end{align}
where $c^{\prime}$ is some constant term independent of the frame parameters. The latter bound stems from Jensen's inequality error estimate bounds that provides an upper-bound for the error of passing the expectation operator from a concave function. The proof steps involves showing that the instantaneous SINR $\gamma(\mx{q},i,\bs{\zeta}(i))$ concentrates around its expectation $\overline{\gamma}$ and then $\overline{\gamma}$ is well approximated by the SINR expression ${\gamma}^{\circ}(\mx{q},i,\mx{W})$.
Define $\mx{B}\triangleq\sum_{\substack{l=1\\l\neq k}}^K \alpha_l^2 \mathcal{A}_l(\mx{z}_l \mx{z}_l^{\rm H})+\bs{\Theta}$, we can write
$\mx{F}_k=\mx{B}+\bs{\Theta}+\rho_{\rm d} \mx{I}_{N_r}$.
\begin{align}\label{eq:t1}
&\tfrac{\overline{\gamma}_k(\mx{q},i,\mx{W})}{N_r}= \tfrac{1}{N_r} \alpha_k^2\left\langle \mx{R}_{\mx{z}_k}, \mx{C}_{\mx{x}_k}\otimes \mathds{E}[{\mx{F}_k}^{-1}]  \right\rangle 
 \stackrel{N_r\rightarrow \infty}{=} \nonumber\\
 &\tfrac{1}{N_r} \alpha_k^2\left\langle \mx{R}_{\mx{z}_k},\mx{C}_{\mx{x}_k}\otimes \bs{\Xi}^{-1}  \right\rangle\triangleq \tfrac{{\gamma}^{\circ}(\mx{q},i,\mx{W})}{N_r}
\end{align}
where $\bs{\Xi}\triangleq \sum_{\substack{l=1\\l\neq k}}^K \tfrac{\alpha_l^2  \mathcal{A}_l(\mx{R}_{\mx{z}_l})}{1+ \alpha_l^2 N_r m_{{\mx{R}_{\mx{z}_l},\mx{B}}}(\rho_{\rm d})}+\bs{\Theta}+\rho_{\rm d}  \mx{I}_{N_r}$ and $m_{{\mx{R}_{\mx{z}_k},\mx{B}}}(\rho_{\rm d})\triangleq\tfrac{1}{N_r}\langle  \mx{R}_{\mx{z}_k}, \mx{C}_{\mx{x}_k}\otimes (\mx{B}+\rho_{\rm d} \mx{I}_{N_r})^{-1} \rangle$ is 
 the Stieltjes transform corresponding to the empirical distribution of eigenvalues of $\mx{B}$.
The approximation error in \eqref{eq:t1} is defined as the difference between $\gamma_k(\mx{q},i,\mx{W})$ and $\gamma_k^{\circ}(\mx{q},i,\mx{W})$ given by:

\begin{align}\label{eq:t3}
   e_N=\tfrac{\alpha_k^2}{N_r}\left\langle \mathcal{A}_k(\mx{R}_{\mx{z}_k}),(\mx{B}+\rho_{\rm d}  \mx{I}_{N_r})^{-1}  \right\rangle-
   \tfrac{\alpha_k^2}{N_r} \left\langle \mathcal{A}_k(\mx{R}_{\mx{z}_k}), \bs{\Xi}^{-1} \right\rangle.\nonumber
\end{align}
It can be shown that $e_N\xrightarrow[]{N_r\rightarrow \infty} 0$ with high probability . Furthermore, the second term above is the Stieltjes transform of a measure and can be found using fixed point iterations \eqref{eq:fixed}.

\bibliographystyle{IEEEtran}
\bibliography{references}

\begin{thebibliography}{10}
\providecommand{\url}[1]{#1}
\csname url@samestyle\endcsname
\providecommand{\newblock}{\relax}
\providecommand{\bibinfo}[2]{#2}
\providecommand{\BIBentrySTDinterwordspacing}{\spaceskip=0pt\relax}
\providecommand{\BIBentryALTinterwordstretchfactor}{4}
\providecommand{\BIBentryALTinterwordspacing}{\spaceskip=\fontdimen2\font plus
\BIBentryALTinterwordstretchfactor\fontdimen3\font minus
  \fontdimen4\font\relax}
\providecommand{\BIBforeignlanguage}[2]{{%
\expandafter\ifx\csname l@#1\endcsname\relax
\typeout{** WARNING: IEEEtran.bst: No hyphenation pattern has been}%
\typeout{** loaded for the language `#1'. Using the pattern for}%
\typeout{** the default language instead.}%
\else
\language=\csname l@#1\endcsname
\fi
#2}}
\providecommand{\BIBdecl}{\relax}
\BIBdecl

\bibitem{rhee2004optimality}
W.~Rhee, W.~Yu, and J.~M. Cioffi, ``The optimality of beamforming in uplink
  multiuser wireless systems,'' \emph{IEEE Transactions on Wireless
  Communications}, vol.~3, no.~1, pp. 86--96, 2004.

\bibitem{rusek2012scaling}
F.~Rusek, D.~Persson, B.~K. Lau, E.~G. Larsson, T.~L. Marzetta, O.~Edfors, and
  F.~Tufvesson, ``Scaling up mimo: Opportunities and challenges with very large
  arrays,'' \emph{IEEE signal processing magazine}, vol.~30, no.~1, pp. 40--60,
  2012.

\bibitem{jafar2001channel}
S.~A. Jafar, S.~Vishwanath, and A.~Goldsmith, ``Channel capacity and
  beamforming for multiple transmit and receive antennas with covariance
  feedback,'' in \emph{ICC 2001. IEEE International Conference on
  Communications. Conference Record (Cat. No. 01CH37240)}, vol.~7.\hskip 1em
  plus 0.5em minus 0.4em\relax IEEE, 2001, pp. 2266--2270.

\bibitem{jafar2004transmitter}
S.~A. Jafar and A.~Goldsmith, ``Transmitter optimization and optimality of
  beamforming for multiple antenna systems,'' \emph{IEEE Transactions on
  Wireless Communications}, vol.~3, no.~4, pp. 1165--1175, 2004.

\bibitem{lu2014overview}
L.~Lu, G.~Y. Li, A.~L. Swindlehurst, A.~Ashikhmin, and R.~Zhang, ``An overview
  of massive mimo: Benefits and challenges,'' \emph{IEEE journal of selected
  topics in signal processing}, vol.~8, no.~5, pp. 742--758, 2014.

\bibitem{soysal2009optimality}
A.~Soysal and S.~Ulukus, ``Optimality of beamforming in fading mimo multiple
  access channels,'' \emph{IEEE transactions on communications}, vol.~57,
  no.~4, pp. 1171--1183, 2009.

\bibitem{hassibi2003much}
``How much training is needed in multiple-antenna wireless links?'' \emph{IEEE
  Transactions on Information Theory}, vol.~49, no.~4, pp. 951--963, 2003.

\bibitem{Fodor:17}
G.~Fodor, N.~Rajatheva, W.~Zirwas, L.~Thiele, M.~Kurras, K.~Guo, A.~Tolli,
  J.~H. Sorensen, and E.~De~Carvalho, ``An overview of massive {MIMO}
  technology components in {METIS},'' \emph{IEEE Communications Magazine},
  vol.~55, no.~6, pp. 155--161, 2017.

\bibitem{Truong:13}
K.~T. {Truong} and R.~W. {Heath}, ``Effects of channel aging in massive {MIMO}
  systems,'' \emph{Journal of Communications and Networks}, vol.~15, no.~4, pp.
  338--351, 2013.

\bibitem{Fodor:23}
S.~Fodor, G.~Fodor, D.~G{\"u}rg{\"u}no{\u{g}}lu, and M.~Telek, ``Optimizing
  pilot spacing in {MU-MIMO} systems operating over aging channels,''
  \emph{IEEE Transactions on Communications}, vol.~71, pp. 3708--3720, 2023.

\bibitem{non_stationary_model}
Z.~Lian, L.~Jiang, C.~He, and D.~He, ``A non-stationary {3-D} wideband {GBSM}
  for {HAP-MIMO} communication systems,'' \emph{IEEE Transactions on Vehicular
  Technology}, vol.~68, no.~2, pp. 1128--1139, 2018.

\bibitem{Fodor:2021}
G.~Fodor, S.~Fodor, and M.~Telek, ``Performance analysis of a linear {MMSE}
  receiver in time-variant rayleigh fading channels,'' \emph{IEEE Transactions
  on Communications}, vol.~69, no.~6, pp. 4098--4112, 2021.

\bibitem{Fodor:22}
G.~{Fodor}, S.{Fodor}, and M.{Telek}, ``{MU-MIMO} receiver design and
  performance analysis in time-varying {R}ayleigh fading,'' \emph{IEEE
  Transactions on Communications}, vol.~70, no.~2, pp. 1214--1228, 2022.

\bibitem{Abeida:10}
H.~Abeida, ``Data-aided {SNR} estimation in time-variant {Rayleigh} fading
  channels,'' \emph{IEEE Transactions on Signal Processing}, vol.~58, no.~11,
  pp. 5496--5507, Nov. 2010.

\bibitem{Hijazi:10}
H.~Hijazi and L.~Ros, ``Joint data {QR}-detection and {Kalman} estimation for
  {OFDM} time-varying {Rayleigh} channel complex gains,'' \emph{IEEE
  Transactions on Communications}, vol.~58, no.~1, pp. 170--177, Jan. 2010.

\bibitem{Kong:2015}
C.~{Kong}, C.~{Zhong}, A.~K. {Papazafeiropoulos}, M.~{Matthaiou}, and
  Z.~{Zhang}, ``Sum-rate and power scaling of massive {MIMO} systems with
  channel aging,'' \emph{IEEE Transactions on Communications}, vol.~63, no.~12,
  pp. 4879--4893, 2015.

\bibitem{Yuan:20}
J.~{Yuan}, H.~Q. {Ngo}, and M.~{Matthaiou}, ``Machine learning-based channel
  prediction in massive {MIMO} with channel aging,'' \emph{IEEE Transactions on
  Wireless Communications}, vol.~19, no.~5, pp. 2960--2973, 2020.

\bibitem{Kim:20}
H.~{Kim}, S.~{Kim}, H.~{Lee}, C.~{Jang}, Y.~{Choi}, and J.~{Choi}, ``Massive
  {MIMO} channel prediction: {Kalman} filtering vs. machine learning,''
  \emph{IEEE Transactions on Communications}, pp. 1--1, 2020, early access.

\bibitem{daei2024towards}
S.~Daei, G.~Fodor, M.~Skoglund, and M.~Telek, ``Towards optimal pilot spacing
  and power control in multi-antenna systems operating over non-stationary
  rician aging channels,'' \emph{arXiv preprint arXiv:2401.13368}, 2024.

\bibitem{Hoydis:13}
J.~Hoydis, S.~T. Brink, and M.~Debbah, ``Massive {MIMO} in the {UL/DL} of
  cellular networks: How many antennas do we need ?'' \emph{IEEE Journal on
  Selected Areas in Communications}, vol.~31, no.~2, pp. 160--171, Feb. 2013.

\bibitem{couillet2011deterministic}
R.~Couillet, M.~Debbah, and J.~W. Silverstein, ``A deterministic equivalent for
  the analysis of correlated {MIMO} multiple access channels,'' \emph{IEEE
  Transactions on Information Theory}, vol.~57, no.~6, pp. 3493--3514, 2011.

\bibitem{hachem2007deterministic}
W.~Hachem, P.~Loubaton, and J.~Najim, ``Deterministic equivalents for certain
  functionals of large random matrices,'' 2007.

\bibitem{bai2010spectral}
Z.~Bai and J.~W. Silverstein, \emph{Spectral analysis of large dimensional
  random matrices}.\hskip 1em plus 0.5em minus 0.4em\relax Springer, 2010,
  vol.~20.

\bibitem{horn2013matrix}
R.~A. Horn and C.~R. Johnson, \emph{Matrix Analysis}, 2nd~ed.\hskip 1em plus
  0.5em minus 0.4em\relax Cambridge University Press, 2013.

\bibitem{jensen_gap}
X.~Gao, M.~Sitharam, and A.~E. Roitberg, ``Bounds on the jensen gap, and
  implications for mean-concentrated distributions,'' \emph{arXiv preprint
  arXiv:1712.05267}, 2017.

\end{thebibliography}

\end{document}